\documentclass[conference]{IEEEtran}
\IEEEoverridecommandlockouts

\usepackage{subcaption}  
\usepackage{color}
\usepackage{pifont}
\usepackage{bbm}
\usepackage{stfloats}
\usepackage[short]{optidef}
\usepackage{multirow}
\usepackage{comment}

\usepackage{amsmath,amssymb,amsthm}
\usepackage{ragged2e} 

\usepackage{array}
\IEEEoverridecommandlockouts

\newtheorem{rem}{Remark}

\usepackage{cite}
\usepackage{amsmath,amssymb,amsfonts}
\usepackage{algorithmic}
\usepackage{graphicx}
\usepackage{textcomp}
\usepackage{xcolor}

\def\BibTeX{{\rm B\kern-.05em{\sc i\kern-.025em b}\kern-.08em
    T\kern-.1667em\lower.7ex\hbox{E}\kern-.125emX}}
\begin{document}
%
\title{Bypassing a Reactive Jammer via NOMA-Based Transmissions in Critical Missions
}

 \author{\IEEEauthorblockN{Mohammadreza Amini, Ghazal Asemian, Michel Kulhandjian, \\ Burak Kantarci, Claude D'Amours and Melike Erol-Kantarci}\\  \vspace{-0.15in}
 \IEEEauthorblockA{School of Electrical Engineering and Computer Science, University of Ottawa, Ottawa, ON, Canada \\
 \texttt{\{gasem093,mkulhand,mamini6,burak.kantarci,cdamours,melike.erolkantarci\}@uottawa.ca}}
 \vspace{-0.35in}
 }

\maketitle

\begin{abstract}
Jamming attacks present a crucial threat to the quality of service in wireless networks, disrupting essential features including reliability, latency, and effective rate specifically in mission-critical applications. This paper introduces and evaluates a NOMA-based model to improve the robustness of wireless networks against jamming attacks. To investigate and address the consequences of a reactive jammer in this context, its effect on substantial network metrics such as reliability, average transmission delay, and the effective sum rate (ESR) under finite blocklength transmissions are mathematically computed, taking by considering the detection probability of the jammer.
 Furthermore, the effect of UEs' allocated power and blocklength on the network metrics is explored.
 Contrary to the existing literature, results show that gNB can mitigate the impact of reactive jamming by decreasing transmit power, making the transmissions covert at the jammer side. Finally, an optimization problem is formulated to maximize the ESR under reliability, delay, and transmit power constraints. It is shown that by adjusting the allocated transmit power to UEs by gNB, the gNB can bypass the jammer effect to fulfill the 0.99999 reliability and the latency of $5ms$ without the need for packet re-transmission.
\end{abstract}


\section{Introduction} \label{Sec.Introduction}
One of the main enabling technologies of the 5G architecture is the Power Domain-Non-Orthogonal Multiple Access (PD-NOMA), 
which enables multiple users to share the same time-frequency resource \cite{islam2016power}. Integrating NOMA into the current paradigm can help increase throughput and reduce transmission latency \cite{ghafoor2022noma}. Moreover, the use of successive interference cancellation (SIC) techniques, enables NOMA to reduce the interference among coexisting users and provides more efficient spectrum utilization \cite{khan2020spectral,saud.2023}. Nonetheless, NOMA systems are vulnerable to jamming attacks due to the sensitivity to received signal power. Disarrangement in the decoding order used by the SIC technique, can result in error in power allocation. On the other hand, decoding the weak user's message by the strong user during SIC introduces vulnerability in the system which should be addressed \cite{akbar2021noma}. 

Among various security vulnerabilities, jamming attacks, which are crucial in mission-critical scenarios such as a military setting are particularly difficult to mitigate. There are different types of jammers capable of targeting a wireless channel. A reactive jammer, 
unlike most types of jammers that continuously transmit their interference signals, 
monitors the communication channel and initiates its attack only when a legitimate transmitter is active \cite{arjoune2020smart}, \cite{vadlamani2016jamming}. 

Mitigating the effect of jamming attacks is of great importance specifically in military communications. In terms of NOMA-based scenarios, there are a few studies that have investigated security problems in response to jamming attacks. 
In \cite{xiao2017reinforcement}, Stackelberg equilibrium is employed to demonstrate the trade-offs experienced by the base station in multiple input multiple output NOMA (MIMO-NOMA) when allocating transmit power to users and meeting the minimum rate requirements of vulnerable users under smart jamming attacks. Q-learning is used to optimize the power allocation strategy by the base station to solve the anti-jamming problem. The authors in \cite{wang2019power} introduce a new framework to reduce the effect of a random jamming attack in MIMO-NOMA transmission using group signal cancellation. To minimize the power consumption considering the required transmission rate, an optimization problem is defined and solved for precoders and MMSE equalizers. The study in \cite{li2019jamming} proposed mitigation of wideband jamming attacks in NOMA by adding a pre-processing tool before SIC and evaluated the results by analyzing BER and spectral efficiency values. The external interference is cancelled through the pre-processing block using a popular blind source separation technique called independent component analysis (ICA), and the internal interference is mitigated using SIC. In \cite{farah2019energy}, the authors have integrated distributed antenna systems (DAS) with NOMA to increase the robustness of the architecture to reactive jamming by increasing the transmit power and the number of active remote radio heads (RRHs). This new scenario determines the resource allocation based on the optimization of energy efficiency. \cite{farah2020efficient} proposed the optimization of RRH and subband selection to mitigate the effect of barrage jammer in the DAS-NOMA system. As the optimal solution depends on the position of users and jammer, the authors suggest an adaptive strategy between three different user-pairing settings of single-SIC NOMA and dual-SIC NOMA with single-antenna transmission, and joint-antenna transmission. To optimize the total transmit power, the study in \cite{tabeshnezhad2023ris} introduced the use of re-configurable intelligent surface (RIS) in the NOMA scenario under the attack of a smart jammer. The fundamental approach taken by the aforementioned research is to increase the transmit power to counteract the jammer while maintaining energy efficiency. However, in a reactive jammer scenario, this may not be effective as the jammer can still sense the transmission and transmit an interfering signal. Apart from the literature and in mitigating the effect of jammer in power-domain \cite{Pirayesh2022}, we focus on reducing the transmit power to deceive the reactive jammer, while maintaining acceptable network performance. Inspired by the covert communication concept \cite{Chen2023}, this work explores bypassing the reactive jammer in a military scenario, taking its detection performance into account.  

To the best of our knowledge, this is the first work that investigates reactive jamming attacks in cluster-based MIMO-NOMA considering the detection probability at the jammer side and the analysis of the network metrics. The communication scheme in our scenario includes short packets and transmission diversity to reflect a military scenario with Ultra-Reliable Low-Latency Communication (URLLC) transmissions. Our main contributions are summarized below.
\begin{itemize}
    \item Derive network metrics such as reliability, average packet delay, and effective sum rate (ESR) in a NOMA-based communication scheme with finite blocklength (FBL) and transmission diversity under the reactive jamming considering the detection probability of the jammer. 

    \item Develop a power-domain jamming mitigation technique in a jamming-aware URLLC NOMA-based scenario by analyzing the effect of gNB transmit power through the derived metrics. 
    \item Formulate an optimization model to obtain the UEs' optimum allocated transmit power alongside other network parameters such as blocklength and the number of packet re-transmissions in a jamming aware scheme.
    
\end{itemize}

 
The paper is organized as follows. In Section \ref{Sec.System_model} the system model is explained. Problem formulation along with derivation of the network metrics in the underlying scenario is presented in Section \ref{Sec.Problem_formulation}. Numerical results and discussions are expressed in Section \ref{Sec.Numerical_result}. Finally, conclusions are given in Section \ref{Sec.Conclusions}. 
\section{System Model} \label{Sec.System_model}
\subsection{Network Model} \label{Sec.System_model.Network}

\begin{figure}
    \centering
    \includegraphics[width=8cm]{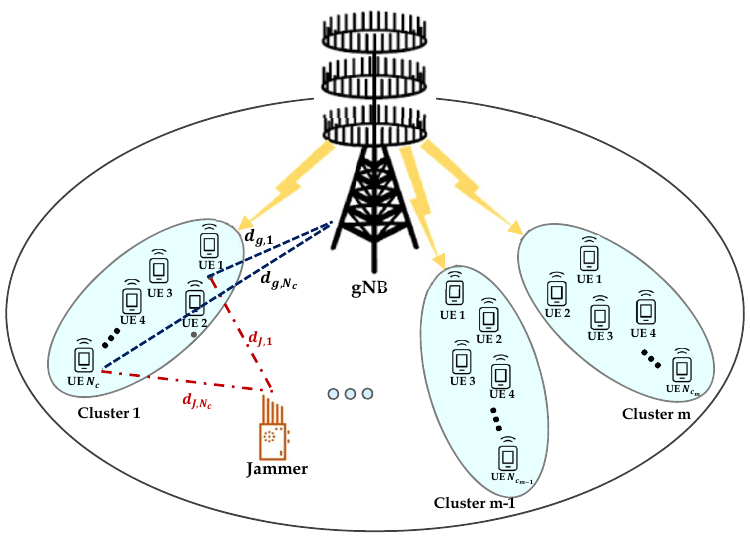}
    \caption{The network structure for cluster-based MIMO-NOMA}
    \label{fig:network} \vspace{-5mm}
\end{figure}

A cluster-based MIMO-NOMA communication scenario with a gNB in the center of the cell and multiple UEs is considered as illustrated in Fig. \ref{fig:network}. The multiple antenna system at the gNB divides the entire cell into $M$ geographical areas called \emph{beam sectors}. Each beam sector forms a NOMA cluster of $N_c$ UEs. All UEs in a NOMA cluster are served by the gNB with the same frequency-time-beam resources, yet at different power levels. The distance from gNB to $\text{UE}_m$ is denoted by $d_{g,m}$ and without loss of generality, UE indices in the NOMA cluster are assumed to be ordered based on their distance, i.e., $d_{g,1}<d_{g,2}<\cdots<d_{g,N_c}$. 

The power allocated to $\text{UE}_m$ by the gNB is denoted by $P_m$ such that $P_t= \sum_{m=1}^{N_c}P_m \leq P_{max}$ in which $P_t$ and $P_{max}$ are the total transmit power and the maximum transmit power of gNB, respectively \footnote{In general, NOMA is significantly sensitive to UEs' allocated power. Hence, determining the channel state of each UE precisely is a challenge, which can be done through signaling.}. The packet arrivals at $\text{UE}_m$ are assumed to follow a Poisson process with the mean arrival rate of $\lambda_m$.

All links experience independent but not necessarily identically distributed Rayleigh block fading, which is assumed to be constant during each transmission time interval (TTI). Given $h_{g,m}$ as the channel coefficient between the gNB and $\text{UE}_m$, the corresponding channel gain $|h_{g,m}|^2$ follows an exponential distribution with mean $d_{g,m}^{-\nu}$, where $\nu$ is the path-loss exponent. Hence, the power of the signal received at $\text{UE}_m$ from gNB at its location is $P_{r,m}=|h_{g,m}|^2 P_m$.
The background noise in all links is modeled by independent and identically distributed (IID) zero-mean additive white Gaussian noise (AWGN) with a variance of $\sigma^2=BN_0$, where $N_0$ and $B$ are the noise spectral density and bandwidth, respectively. 
\graphicspath{{figs/}}
\vspace{2mm}

\subsection{Jammer Model} \label{Sec.System_model.Jamming}
A reactive jammer is assumed such that it transmits power whenever it senses any transmission in the channel\footnote{We assume that the jamming signal follows a Gaussian distribution.}. In particular, the jammer measures the energy of the intended spectrum and jams only if the corresponding test statistic, $\Xi$, is higher than a predefined threshold. Let $P_J$ and $P_{th}$ be the jammer's transmit power and its triggering threshold, respectively. The distance from gNB to the jammer and from the jammer to $\text{UE}_m$ is $d_{g,J}$ and $d_{J,m}$, respectively. The former is assumed to be known at gNB\footnote{Note that the distance between the jammer and gNB is assumed to be known as a result of jammer localization in a jamming-aware scenario.}. Due to the uncertainty of the communication channel, the detection process at the jammer side is imperfect. The jammer's detection can be modeled by two error probabilities: false-alarm $\mathbb{P}_f$ and miss-detection $\mathbb{P}_{md}$. The detection process at the jammer can be viewed as a hypothesis test problem with $H_0$ and $H_1$ hypotheses as the absence and presence of gNB's signal, respectively. In the absence of any information from gNB, both gNB signal and noise are assumed Circularly Symmetric Complex
Gaussian (CSCG). Without considering the fading effect, the detection probability, $\mathbb{P}_d=1-\mathbb{P}_{md}$, is derived as \cite{Liang2008}, \vspace{-5mm}

	\begin{equation}\label{Eq.P_d}
		\begin{split}
			\mathbb{P}_d=\mathcal{Q}\left(\frac{P_{th}-(\gamma_J+1)\sigma^2}{\sqrt{\frac{1}{N} \left(\gamma_J+1\right)^2\sigma^4}}\right) \, ,
		\end{split} 
	\end{equation}
where $\mathcal{Q}$ is complementary distribution function of the
standard Gaussian and $\gamma_{J}=\frac{|h_{g,J}|^2 P_t}{\sigma^2}$ is the Signal-to-Noise Ratio (SNR) at the jammer side. Moreover, $N$ is the number of samples taken for the detection process. Under Rayleigh fading channel, detection probability ($\mathbb{P}_d^{Ray}$) can be written as,\vspace{-4mm}

\begin{equation}\label{Eq.P_d_ray}
		\begin{split}
			\mathbb{P}_d^{Ray}=\int_0^{+\infty} \mathbb{P}_d f_{\gamma_{_J}}(\gamma)d\gamma \, ,
		\end{split} 
	\end{equation} where $f_{\gamma_{_J}}(\gamma)=\frac{\sigma^2 \, d_{g,J}^\nu}{P_t} \, exp\left( {\frac{-\sigma^2 \, d_{g,J}^\nu}{P_t}\gamma}\right)$, is the PDF of $\gamma_{_J}$.

\subsection{Transmission Model} \label{Sec.System_model.transmission}
To achieve URLLC, short packets with transmission diversity are assumed in the transmission protocol with $L$ number of successive re-transmissions \cite{Shirvanimoghaddam_URLCC1, Lee}.  
It is assumed that UEs transmit their data packets in FBL regime to achieve low packet latency. However, in such a case, Shannon's capacity is no longer applicable since the decoding block error cannot be ignored. Thus, given a blocklength of $n_b$ for $n_d$ data bits per data packet, the instantaneous block error rate (BLER) of decoding $\text{UE}_i$'s signal at the location of $\text{UE}_m$ is approximated as \cite{Polyanskiy2010} \vspace{-3mm}, 

\begin{equation}\label{Eq.FBL_error}
	\epsilon_{i,m} =  \mathcal{Q}\Bigg(\sqrt{\frac{n_b}{\chi\left(\gamma_{i,m }\right)}} \bigg(\mathcal{C}\big(\gamma_{i,m }\big)-\frac{n_d}{n_b}\bigg)\Bigg),
\end{equation}
where $\mathcal{C}\left(\gamma_{i,m } \right) =  \log_2(1 + \gamma_{i,m})$ is the Shannon capacity of $\text{UE}_i$, while $\chi\left(\gamma_{i,m }\right) = \left(1 - \frac{1}{1 + \gamma^2_{i,m }}\right)\left(\log_2e\right)^2$ represents the channel dispersion. Furthermore, $\gamma_{i,m}$ is the received Signal-to-Interference plus Noise Ratio (SINR) of $\text{UE}_i$'s signal at the gNB which is given as (\ref{Eq.gamma_i}) under non-jammed transmissions\footnote{Successive Interference Cancellation (SIC) is assumed at the gNB.}.\vspace{-3mm}

\begin{equation}\label{Eq.gamma_i}
	\gamma_{i,m }= 
     \frac{|h_{g,m}|^2 P_i}{\sum_{l=1}^{i-1}|h_{g,m}|^2 P_l+\sigma^2} \quad i \geq m
\end{equation}

For the jammed links, i.e., $\Xi>P_{th}$, the BLER is denoted by $\epsilon_{i,m}^J$ which follows the same expression as  (\ref{Eq.FBL_error}) but calculated based on Signal-to-Interference plus Noise and Jamming Ratio (SINJR) $\gamma_{i,m}^{J}$ as, \vspace{-7mm}

\begin{equation}\label{Eq.gamma_i_J}
	\gamma_{i,m}^{J}= \frac{|h_{g,m}|^2 P_i}{\sum_{l=1}^{i-1}|h_{g,m}|^2 P_l+ |h_{J,m}|^2 P_J +\sigma^2 }   \quad i \geq m \, ,
\end{equation} where $h_{J,m}$ is the channel coefficient between the jammer and $\text{UE}_m$. 
A typical frame herein consists of two parts, namely, header (contains information related to the payload) and payload. Therefore, the whole frame duration is $T_f=T_h +T_p$ where $T_h$ and $T_p$ are the header and payload duration, respectively where $T_p=\frac{n_b}{B}$ due to the short packet transmissions.

\section{Derivation of the Network Metrics in a Jamming-Aware Scenario}
\label{Sec.Problem_formulation}
We consider the detection probability at the jammer side in the derivations with the aim of bypassing the jammer, and formulate the optimization problem to reach the URLL requirements while mitigating the effect of the jammer. 

Any detection process that comes with uncertainty, such as detecting gNB's signal by the jammer, is non-deterministic. 
On the other hand, all the network metrics are affected by such detection. Since NOMA-based transmissions are considered herein, in case the jammer detects any transmission, the effective rate, reliability, and packet latency are significantly affected since a NOMA receiver is more vulnerable to the received power than OMA due to SIC. Below network metrics 
 for the underlying NOMA-based scenario 
are derived under the reactive jamming mitigation approach.

\subsubsection{Reliability} 

Reliability of $\text{UE}_m$, $\mathcal{R}_m$, is the probability that a typical transmitted packet by gNB is successfully decoded at the $\text{UE}_m$ location. To calculate the $\mathcal{R}_m$ under $H_1$, we first find the probability of successfully decoding a $\text{UE}_m$'s packet without considering its retransmissions, i.e., $\mathcal{P}_m$. Conditioning on the test statistic value calculated by the jammer, $\mathcal{P}_m$ can be written as,\vspace{-5mm}

\begin{equation}\label{Eq.P1}
\begin{split}
	\mathcal{P}_m&= \mathcal{P}_{m \mid \Xi>P_{th}} \, \textit{Pr}(\Xi>P_{th} ) \\
 & \hspace{5mm} + \> \mathcal{P}_{m \mid \Xi<P_{th}} \, \textit{Pr}(\Xi<P_{th} )\\
 &= p_m^J \mathbb{P}_d^{Ray} +p_m (1-\mathbb{P}_d^{Ray}) \,  ,
 \end{split}
\end{equation} where $p_m^J$ and $p_m$ denote the probability of successfully decoding $\text{UE}_m$'s packet with and without jammer transmission, respectively. Since SIC is used, $\text{UE}_m$ should decode the signals intended for the $\text{UE}_m$ for all $i < m$ before decoding its own signals. Hence, $p_m^J$ and $p_m$ are obtained by (\ref{Eq.P2_J}) and (\ref{Eq.P2}).\vspace{-2mm}
\begin{equation}\label{Eq.P2_J}
	p_m^J=\prod_{i=m}^{N_c} (1-\epsilon_{i,m}^J) \quad 
\end{equation}
\vspace{-3mm}
\begin{equation}\label{Eq.P2} 
	p_m=\prod_{i=m}^{N_c} (1-\epsilon_{i,m}) \quad 
\end{equation}
Considering all the packet replicas due to transmission diversity, $\mathcal{R}_m$ can be derived using binomial expansion as,\vspace{-5mm}

\begin{equation}\label{Eq.Rel}
\begin{split}
	\mathcal{R}_m &=\sum_{k=1}^L \binom{L}{k} 
\mathcal{P}_m^{k} \left(1-\mathcal{P}_m \right)^{L-k} \\
&=\sum_{k=0}^L \binom{L}{k} 
\mathcal{P}_m^{k} \left(1-\mathcal{P}_m \right)^{L-k}- \left(1-\mathcal{P}_m \right)^{L} \\
&=  1-\left(1-\mathcal{P}_m \right)^{L}
\end{split}
\end{equation}

\subsubsection{Packet Delay} 
The average transmission delay of a typical packet for $\text{UE}_m$, $\mathcal{\overline{D}}_m$, is the sum of two main components: average packet transmission time (on air latency) and average queuing delay (i.e., the gNB buffer waiting time). When viewed at the gNB side and transmission diversity, the underlying packet transmission scheme can be modeled as an M/D/1 queuing system having a Poisson packet arrival process with the mean arrival rate of $\lambda_m$ and constant service rate of $1/(L\times T_f)$. Therefore, the average transmission delay is\footnote{Eq. (10) holds true as long as the queue is stable.} \cite{Bhat2015},\vspace{-5mm}

\begin{equation}\label{Eq.delay}
    \begin{split}
    	\mathcal{\overline{D}}_m=\frac{2-\lambda_m L T_f}{2(1-\lambda_mLT_f)} LT_f \, .
     \end{split}
\end{equation} 

\subsubsection{Effective Sum Rate} 

Effective Sum Rate, $\eta$, is defined as the sum of the effective rates for all UEs in the NOMA cluster, and the effective rate for each user- say $\text{UE}_m$, $r_m$- is defined as the number of bits delivered successfully over time unit. Let $\mathcal{B}_m$ be the effective bits (error-free decoded bits) delivered to $\text{UE}_m$ per a typical transmitted packet. Hence, the average effective rate of $\text{UE}_m$ is derived as,\vspace{-3mm}

\begin{equation}\label{Eq.user_rate}
    \begin{split}
    	r_m&= \frac{\mathbb{E} \left[ \mathcal{B}_m\right]}{\mathcal{\overline{D}}_m} =\frac{n_d \times \mathcal{R}_m}{\mathcal{\overline{D}}_m}\\
     &= \frac{2n_d (1-\lambda_mLT_f) \left(1-\left(1-\mathcal{P}_m \right)^{L}\right)}{LT_f(2-\lambda_mLT_f)} \, ,
     \end{split}
\end{equation} Note that since $n_b$ contains redundant bits related to coding schemes, $n_d$ is used as the number of information bits that are sent by gNB to a typical UE. Therefore, the effective number of (information) bits equals $n_d \times \mathcal{R}_m$. Finally, the ESR is calculated as $\eta=\sum_{m=1}^{N_c}r_m$.

The obtained analytical derivations can be utilized at the gNB to derive the optimum power allocated to each UE. The optimization problem can be formulated to maximize the effective sum rate in a NOMA-based reactive jammer scenario subject to reliability, packet delay, SIC decoding order, and transmit power limit as shown below:\vspace{-3mm}

\begin{equation*}
\hspace{-55mm}\underline{\text{\textbf{ESR-Max:}}}
\end{equation*}\vspace{-0.325in}
\begin{maxi!}[2]
	{n_b, L,\mathbf{P}}{\eta=\sum_{m=1}^{N_c}r_m\label{Eq.maxa}}{}{}
	\addConstraint{\mathcal{R}_{m}}{\geq \delta_{th}^{r}, \qquad \, \, \, \forall m \in \{1,..., N_c\} \label{Eq.maxb}}
	\addConstraint{\mathcal{\overline{D}}_m }{\leq \delta_{th}^{d}, \qquad \, \, \, \forall m \in \{1,..., N_c\}\label{Eq.maxc}}
    \addConstraint{0< P_{m}}{\leq P_{m+1} \, \, \, \forall m \in \{1,..., N_c-1\} \label{Eq.maxd}}
	\addConstraint{\sum_{m=1}^{N_c} P_m \leq P_{max} \label{Eq.maxe}}
	\addConstraint{n_b, L}{\in \{1, 2, ...\} .\label{Eq.maxf}}
\end{maxi!}
where $\mathbf{P} = \left[P_{1}, P_{2}, ..., P_{N_c}  \right]$ is the transmit power vector of all UEs, and L is the number of successive retransmissions (L=1, means no retransmission). Constraint (\ref{Eq.maxb}) ensures that the minimum reliability $\delta_{th}^r$ is satisfied, while Constraint (\ref{Eq.maxc}) is the latency requirement, which is at most $\delta_{th}^d$. Constraint (\ref{Eq.maxd}) and (\ref{Eq.maxe})
enforces the SIC decoding order and maximum gNB transmit power. Finally, (\ref{Eq.maxf}) defines the range of values the decision variables take.

\begin{rem}\label{optimization_prob}
	Problem \textbf{ESR-Max} is classified as a mixed-integer non-linear programming (MINLP) problem, which is non-convex and NP-hard \cite{G_Nemhauser88}. This can be verified from the derived non-linear expressions of $\mathcal{R}_{m}$, $\mathcal{\overline{D}}_m$, $\eta$, and the integer-valued decision variables $n_b$ and $L$. 

Therefore, to solve the optimization problem, a genetic algorithm is used in the numerical results.
\end{rem}

\section{Numerical Results} \label{Sec.Numerical_result}
\subsection{Parameter Settings}

We consider the NOMA cluster size of two UEs\footnote{The two-node NOMA communication is an elementary block of NOMA, addressed in the 3GPP \cite{3GPPTR36.859}.}, i.e., $N_c=2$. Subcarrier spacing of $60\text{kHz}$ which provides mini-slots for URLLC is adopted as 5G numerology. Table \ref{Table1.Param} summarizes the simulation settings adopted from \cite{Iwabuchi,Le2021}.

\begin{table}[!ht]
	\centering
	\caption{Simulation Parameters } \label{Table1.Param}
	\begin{tabular}{c|c||c|c} \hline
		\textbf{Parameter}	&	\textbf{Value}  &	\textbf{Parameter}  &  \textbf{Value}  \\
		\hline \hline
		$B$ & $720 \, \text{kHz}$ & $N_c$ & $2$ \\ \hline
		$\nu$ & $2.5$ & $T_{h}$ & $1 \,  \text{ms}$ \\  \hline
		$d_{g,1},d_{g,2}$ & $10,50 \, m$ & $N_0$ & $-174 \, \text{dBm/Hz} $ \\ \hline	
		$d_{g,J}$ & $30 \,m$  & $n_d$ &  $32 \, \text{Bytes} $ \\ \hline
		$d_{J,1}, d_{J,2}$ & $27.3, 77 \, m$ & $P_J$ & $20\, \textit{mW}$ \\ \hline
	\end{tabular}
\end{table}
\subsection{Performance Results}
\begin{figure}[htp]
    \centering
    \includegraphics[width=.8\linewidth]{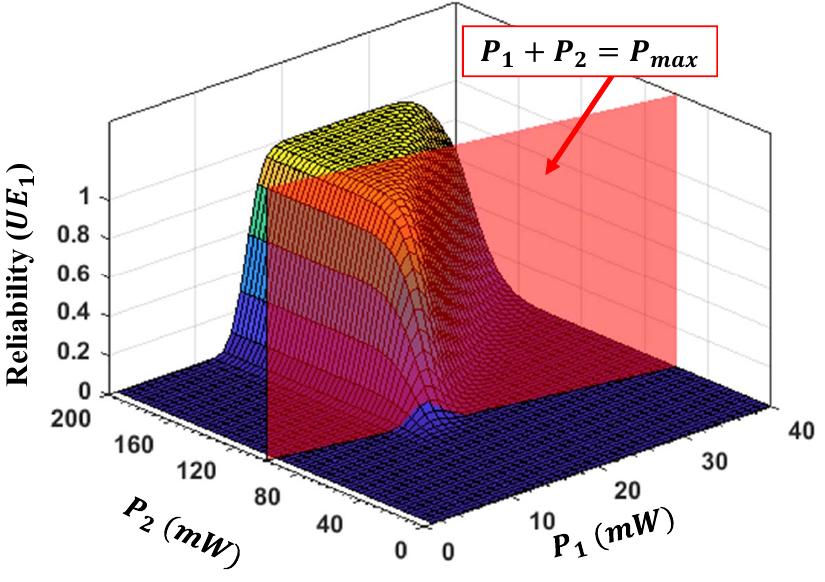}
    \caption{$UE_1$ Reliability vs. transmit power $P_1$ and $P_2$ under barrage jamming. $L=2$, $n_b=80$.}
    \label{fig.Rel_power_BJ} \vspace{-5mm}
\end{figure}

To understand the reactive jamming scenario, we first plot the reliability of $\text{UE}_1$ under barrage jammer, which continuously injects its power into the communication link, in Fig. \ref{fig.Rel_power_BJ}.Note that due to the lack of space, we only focus on the $\text{UE}_1$ metrics. However, $\text{UE}_2$ experiences the same trend. The gNB transmit power plane, i.e., $P_1+P_2 = P_{max}$ is also included in the figure. The area in front of this plane is the feasible region that passes the gNB transmit power constraint which is $P_1+P_2 \leq P_{max}$. As seen, the solution space that meets the reliability requirement of \textit{five-nines}, i.e., $\mathbf{R}_i \geq 0.99999$, does not fulfill the transmit power constraint. Therefore, the link is completely jammed under the barrage jammer. 

\graphicspath{{Figs/}}

\vspace{2mm}

\begin{figure}[htp]
    \centering
    \includegraphics[width=.9\linewidth] {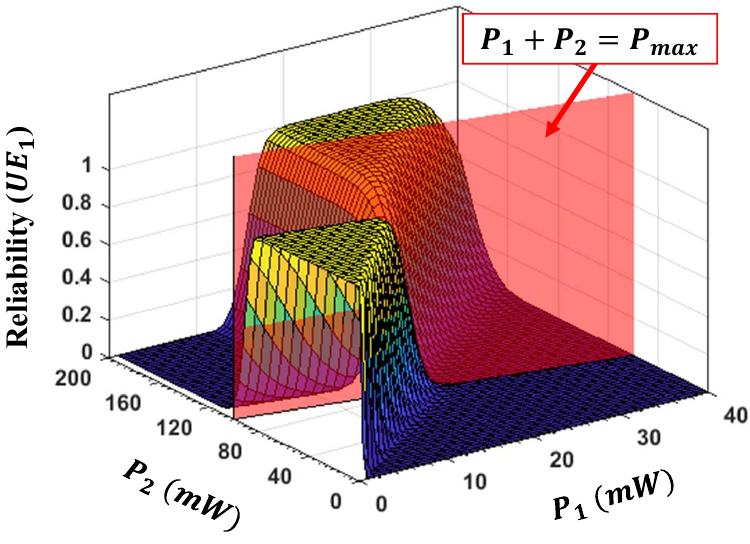}
    \caption{$UE_1$ reliability vs. transmit power $P_1$ and $P_2$ under reactive jamming. $L=2$, $n_b=80$.}
    \label{fig.Rel_power_RJ} \vspace{-5mm}
\end{figure}

Fig. \ref{fig.Rel_power_RJ} presents the $\text{UE}_1$'s reliability under reactive jamming when the gNB is aware of the jammer. Interestingly, the region $P_1+P_2 \leq P_{max}$ contains some solutions for the transmit power. We show that the gNB can reduce its power to mitigate the effect of the jammer in order to meet both the reliability requirement and transmit power constraint. Without considering the detection probability of the jammer, the gNB tries to increase its power due to the power control mechanism. This does not help mitigate the effect of the jammer as it reaches the maximum transmit power.

\begin{figure}[htp]
    \centering
    \includegraphics[width=.7\linewidth]{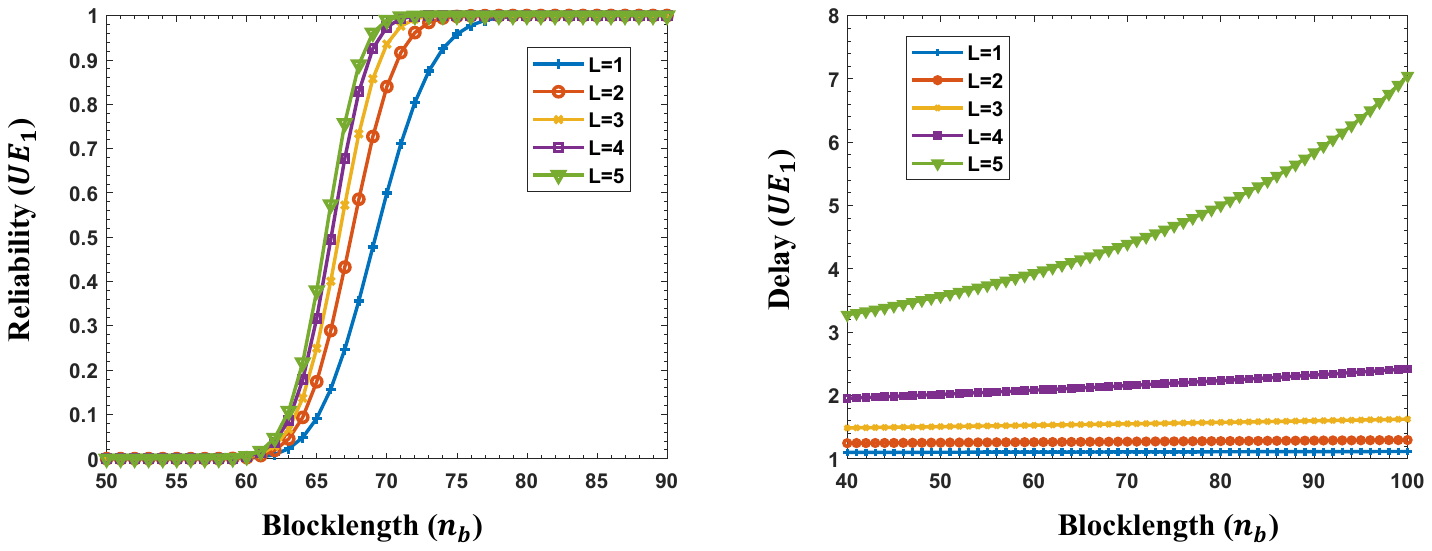}
    \caption{$UE_1$ reliability-blocklength. \{$P_1, P_2$\}=\{10, 60\}[mW]}
    \label{fig.Rel_nb} \vspace{-2mm}
\end{figure}

\begin{figure}[htp]
    \centering
    \includegraphics[width=.7\linewidth]{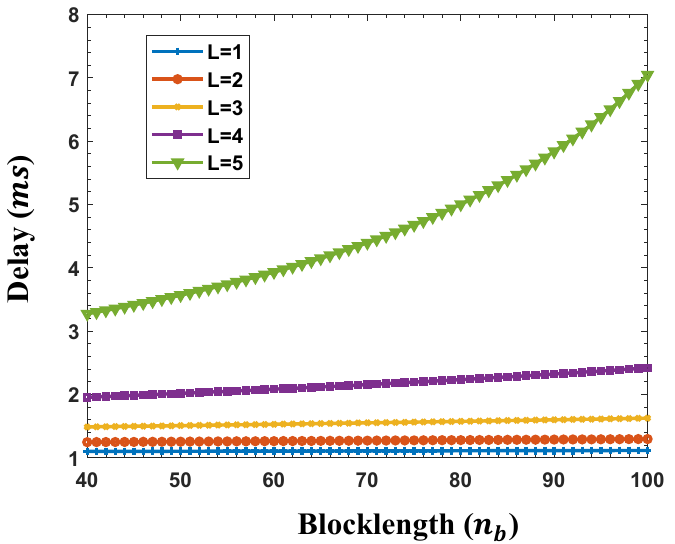}
    \caption{Avg. trans. delay-blocklength\{$P_1, P_2$\}=\{10, 60\}[mW]}
    \label{fig.Delay_nb} \vspace{-7mm}
\end{figure}

\graphicspath{{Figs/}}
\vspace{2mm}
The effect of the blocklength on $\text{UE}_1$'s reliability is depicted in Fig. \ref{fig.Rel_nb} for various number of packet re-transmissions. It is observed that the reliability experiences a sharp increase for $60<n_b<70$ and then it reaches to its maximum value. This increase in reliability is due to the fact that the decoding error probability decreases with an increase in blocklength and then, it reaches to the maximum of $1$ when increasing the blocklength no longer helps improve the reliability. It is observed that for $n_b < 79$ the reliability is lower than the prescribed target of five-nines when $L=1$, i.e., without any re-transmission. However, by increasing the number of re-transmissions, the reliability tends to meet the target in shorter blocklengths. For example, when $L=5$, the reliability reaches $100 \%$ for $n_b >68$. On the other hand, increasing the number of re-transmissions is not always the choice as it threatens the delay requirement in URLLC applications which is depicted in Fig. \ref{fig.Delay_nb}. The higher the $n_b$ and $L$ are, the greater the transmission delay. Therefore, an optimum choice for such parameters must be made through the optimization Problem \textbf{ESR-Max} in (12). 

\graphicspath{{Figs/}}
\vspace{2mm}
Note that increasing $n_b$ increases the redundancy and lengthening the frame that is reflected as a higher delay in Fig. \ref{fig.Delay_nb}. Furthermore, as $L$ increases, the transmission delay increases significantly which results from the homographic behavior of (\ref{Eq.delay}) as it is clearly seen for $L=5$.

\begin{figure}[htp]
    \centering
    \includegraphics[width=.8\linewidth]{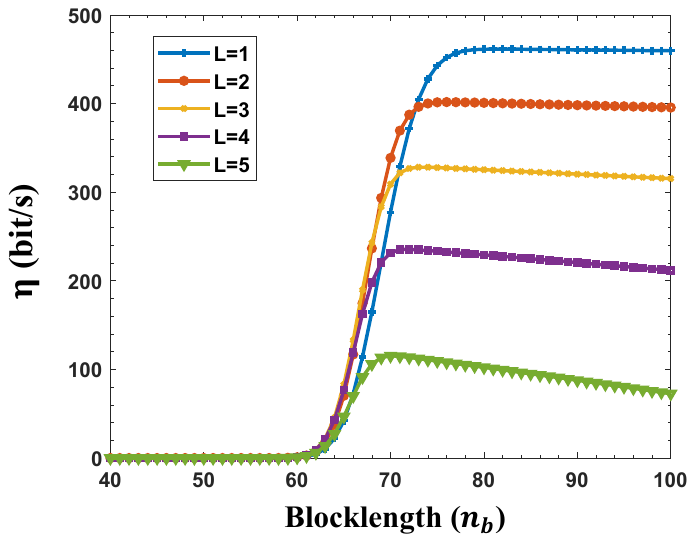}
    \caption{ESR vs. blocklength \{$P_1, P_2$\}=\{10, 60\}[mW]}.
    \label{fig.ESR_nb} \vspace{-8mm}
\end{figure}

\graphicspath{{Figs/}}
\vspace{2mm}
Fig. \ref{fig.ESR_nb} explores the ESR as a function blocklength under various numbers of packet re-transmissions. As can be observed, ESR experiences a sharp increase for $60<n_b<70$ and then tends to decrease slowly. This is because the reliability increases sharply when blocklength increases (as a result of reducing the decoding error rate) and hence, the number of error-free decoded blocks increases. However, when the reliability reaches its maximum value, increasing the blocklength only increases the frame length. Hence, the number of effective bits per time unit decreases, causing the ESR to gently decreases. Furthermore, the higher the number of re-transmissions is, the lower the ESR, i.e., the expense of ultra-reliable transmissions.

\graphicspath{{Figs/}}
\vspace{2mm}
\subsection{Optimizing the Network Parameters}

To solve the problem \textbf{ESR-Max}, the genetic algorithm (GA) is used. The optimization problem in (12) is solved for $N_c=2$, $\delta_{th}^r=0.99999$, $\delta_{th}^d=5\times 10^{-3} ms$, and $P_{max}=100 mW$. Furthermore, the objective function in the GA is defined as $ Obj=-\eta=-\sum_{m=1}^{N_c=2} r_m$.  The GA settings are listed in Table \ref{Table1.GA_Param}. The tolerance for the constraints is set according to the most stringent constraint. Since the reliability threshold $\delta_{th}^r=1-10^{-5}$, $ConstraintTolerance$ is set to $10^{-6}$ to ensure it is not violated. The other parameters are set empirically and based on initial executions.

\begin{figure}[htp]
    \centering
    \includegraphics[width=0.95\linewidth]{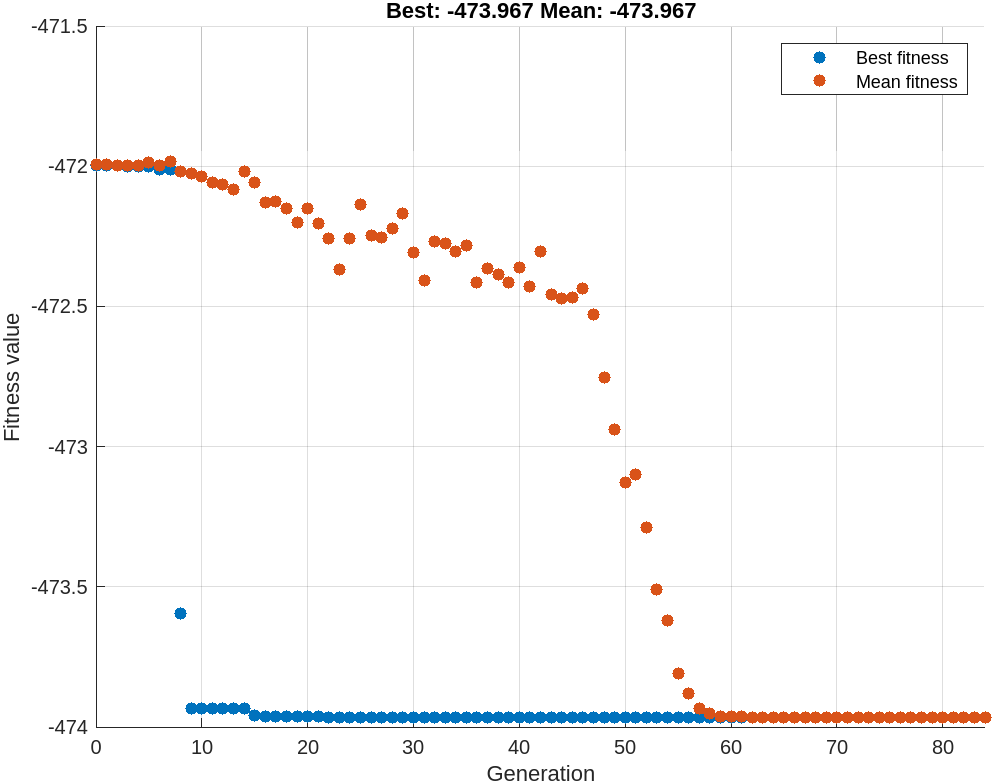}
    \caption{Genetic Algorithm convergence process.}
    \label{fig.GA} \vspace{-3mm}
\end{figure}

Fig. \ref{fig.GA} shows the fitness value of the objective function versus the number of generations. The algorithm converges in less than $60$ generations. The optimum obtained values for the decision variables are $(P_1,P_2,L,n_b)=(13,57,1,83)$, meaning that by allocating the transmit power of $13$ and $57$ mW to $\text{UE}_1$ and $\text{UE}_2$, respectively with blocklength of $83$ without any packet re-transmissions, the gNB can bypass the jammer and meets the defined latency-reliability requirements. Note that since the reliability of five-nines with $L=1$ and $n_b=83$ is met, there is no need for packet re-transmission. However, based on the jammer's configuration (location and transmit power the optimum values might change.
\begin{table}[!ht]
	\centering
	\caption{Genetic Algorithm Parameters} \label{Table1.GA_Param}
	\begin{tabular}{c|c||c|c} \hline
		\textbf{Parameter}	&	\textbf{Value}  &	\textbf{Parameter}  &  \textbf{Value}  \\
		\hline \hline
		$PopulationSize$ & $150 $ & $ConstraintTolerance$ & $10^{-6}$ \\ \hline
		$MaxGenerations$ & $200$ & $CrossoverFraction$ & $0.8$ \\  \hline
		$FunctionTolerance$ & $10^{-10}$ & $MaxStallGenerations$ & $50$ \\ \hline	
	\end{tabular} 
\end{table}

\graphicspath{{Figs/}}
\vspace{2mm}

\section{Conclusions} \label{Sec.Conclusions}
\vspace{-0mm}
This paper has investigated the dynamics of network metrics in the presence of a reactive jammer. A cluster-based MIMO-NOMA network with short packet and diversity transmissions has been considered for mission-critical applications in a military scenario, and the reliability, packet delay, and effective sum rate have been derived for further investigation under reactive jamming attack. Under a jamming-aware scenario where the gNB is aware of the presence of the jammer, we have shown that the gNB adapts its transmit power to be miss-detected by the jammer, taking the jammer detection probability into account. Finally, an optimization model has been formulated based on the derived network metrics and solved by the genetic algorithm for the network parameters in order to maximize the effective sum rate under URLL constraints. It is shown that by adjusting the allocated transmit power to UEs by gNB, the gNB can bypass the jammer effect to meet the 0.99999 reliability target and achieve a $5ms$ latency without requiring packet re-transmission.  

\section*{Acknowledgement}
\vspace{-0.08in}

This work was supported in part by funding from the Innovation for Defence Excellence and Security (IDEaS) program from the Department of National Defence (DND) and in part by the NSERC CREATE TRAVERSAL Program.


\bibliographystyle{IEEEtran}

\begin{thebibliography}{10}
\vspace{-0.05in}
\providecommand{\url}[1]{#1}
\csname url@samestyle\endcsname
\providecommand{\newblock}{\relax}
\providecommand{\bibinfo}[2]{#2}
\providecommand{\BIBentrySTDinterwordspacing}{\spaceskip=0pt\relax}
\providecommand{\BIBentryALTinterwordstretchfactor}{4}
\providecommand{\BIBentryALTinterwordspacing}{\spaceskip=\fontdimen2\font plus
\BIBentryALTinterwordstretchfactor\fontdimen3\font minus
  \fontdimen4\font\relax}
\providecommand{\BIBforeignlanguage}[2]{{%
\expandafter\ifx\csname l@#1\endcsname\relax
\typeout{** WARNING: IEEEtran.bst: No hyphenation pattern has been}%
\typeout{** loaded for the language `#1'. Using the pattern for}%
\typeout{** the default language instead.}%
\else
\language=\csname l@#1\endcsname
\fi
#2}}
\providecommand{\BIBdecl}{\relax}
\BIBdecl

\bibitem{islam2016power}
S.~R. Islam, N.~Avazov, O.~A. Dobre, and K.-S. Kwak, ``Power-domain
  non-orthogonal multiple access (noma) in 5g systems: Potentials and
  challenges,'' \emph{IEEE COMST}, vol. 19/2, pp. 721--742, 2016.

\bibitem{ghafoor2022noma}
U.~Ghafoor, M.~Ali, H.~Z. Khan, A.~M. Siddiqui, and M.~Naeem, ``Noma and future
  5g \& b5g wireless networks: A paradigm,'' \emph{Journal of Network and
  Computer Applications}, vol. 204, p. 103413, 2022.

\bibitem{khan2020spectral}
W.~U. Khan, J.~Liu, F.~Jameel, V.~Sharma, R.~J{\"a}ntti, and Z.~Han, ``Spectral
  efficiency optimization for next generation noma-enabled iot networks,''
  \emph{IEEE Trans on Vehicular Tech.}, vol. 69/12, pp. 15\,284--15\,297, 2020.

\bibitem{saud.2023}
S.~Althunibat, H.~Hassan, T.~Khattab, and N.~Zorba, ``A new {NOMA}-based
  two-way relaying scheme,'' \emph{IEEE Transactions on Vehicular Technology},
  vol.~72, no.~9, pp. 12\,300--12\,310, 2023.

\bibitem{akbar2021noma}
A.~Akbar, S.~Jangsher, and F.~A. Bhatti, ``Noma and 5g emerging technologies: A
  survey on issues and solution techniques,'' \emph{Computer Networks}, vol.
  190, p. 107950, 2021.

\bibitem{arjoune2020smart}
Y.~Arjoune and S.~Faruque, ``Smart jamming attacks in 5g new radio: A review,''
  in \emph{IEEE Annual Computing and Comm. Workshop \& conf.}, 2020, pp.
  1010--1015.

\bibitem{vadlamani2016jamming}
S.~Vadlamani, B.~Eksioglu, H.~Medal, and A.~Nandi, ``Jamming attacks on
  wireless networks: A taxonomic survey,'' \emph{International Journal of
  Production Economics}, vol. 172, pp. 76--94, 2016.

\bibitem{xiao2017reinforcement}
L.~Xiao, Y.~Li, C.~Dai, H.~Dai, and H.~V. Poor, ``Reinforcement learning-based
  noma power allocation in the presence of smart jamming,'' \emph{IEEE Trans on
  Vehicular Tech.}, vol. 67/4, pp. 3377--3389, 2017.

\bibitem{wang2019power}
H.~Wang, Y.~Fu, R.~Song, Z.~Shi, and X.~Sun, ``Power minimization precoding in
  uplink multi-antenna noma systems with jamming,'' \emph{IEEE Trans on Green
  Communications and Netw.}, vol. 3/3, pp. 591--602, 2019.

\bibitem{li2019jamming}
J.~Li and M.~Fan, ``Jamming suppression in downlink noma using independent
  component analysis,'' in \emph{Int Conf on Comm. Tech}, 2019, pp. 164--168.

\bibitem{farah2019energy}
J.~Farah, J.~Akiki, and E.~P. Simon, ``Energy-efficient techniques for
  combating the influence of reactive jamming using non-orthogonal multiple
  access and distributed antenna systems,'' in \emph{2019 Wireless
  Telecommunications Symposium (WTS)}.\hskip 1em plus 0.5em minus 0.4em\relax
  IEEE, 2019, pp. 1--7.

\bibitem{farah2020efficient}
J.~Farah, E.~P. Simon, P.~Laly, and G.~Delbarre, ``Efficient combinations of
  noma with distributed antenna systems based on channel measurements for
  mitigating jamming attacks,'' \emph{IEEE Systems Journal}, vol.~15, no.~2,
  pp. 2212--2221, 2020.

\bibitem{tabeshnezhad2023ris}
A.~Tabeshnezhad, A.~L. Swindlehurst, and T.~Svensson, ``Ris-assisted
  interference mitigation for uplink noma,'' in \emph{IEEE Wireless
  Communications and Networking Conference}.\hskip 1em plus 0.5em minus
  0.4em\relax IEEE, 2023, pp. 1--5.

\bibitem{Pirayesh2022}
H.~Pirayesh and H.~Zeng, ``Jamming attacks and anti-jamming strategies in
  wireless networks: A comprehensive survey,'' \emph{IEEE COMST}, vol.~24,
  no.~2, pp. 767--809, 2022.

\bibitem{Chen2023}
X.~Chen, J.~An, Z.~Xiong, C.~Xing, N.~Zhao, F.~R. Yu, and A.~Nallanathan,
  ``Covert communications: A comprehensive survey,'' \emph{IEEE Communications
  Surv. \& Tutorials}, vol.~25, no.~2, pp. 1173--1198, 2023.

\bibitem{Liang2008}
Y.-C. Liang, Y.~Zeng, E.~C. Peh, and A.~T. Hoang, ``Sensing-throughput tradeoff
  for cognitive radio networks,'' \emph{IEEE Transactions on Wireless
  Communications}, vol.~7, no.~4, pp. 1326--1337, 2008.

\bibitem{Shirvanimoghaddam_URLCC1}
M.~Shirvanimoghaddam, M.~S. Mohammadi, R.~Abbas, A.~Minja, C.~Yue, B.~Matuz,
  G.~Han, Z.~Lin, W.~Liu, Y.~Li, S.~Johnson, and B.~Vucetic, ``Short
  block-length codes for ultra-reliable low latency communications,''
  \emph{IEEE Communications Magazine}, vol.~57, no.~2, pp. 130--137, 2019.

\bibitem{Lee}
H.~Lee and Y.-C. Ko, ``Physical layer enhancements for ultra-reliable
  low-latency communications in 5g new radio systems,'' \emph{IEEE
  Communications Standards Magazine}, vol.~5, no.~4, pp. 112--122, 2021.

\bibitem{Polyanskiy2010}
Y.~{Polyanskiy}, H.~V. {Poor}, and S.~{Verdu}, ``Channel coding rate in the
  finite blocklength regime,'' \emph{IEEE Trans. Inf. Theory}, vol.~56, no.~5,
  pp. 2307--2359, May 2010.

\bibitem{Bhat2015}
U.~N. Bhat, \emph{An Introduction to Queueing Theory, Modeling and Analysis in
  Applications.}\hskip 1em plus 0.5em minus 0.4em\relax Birkhäuser Basel,
  2015.

\bibitem{G_Nemhauser88}
G.~Nemhauser, \emph{Integer and Combinatorial Optimization}.\hskip 1em plus
  0.5em minus 0.4em\relax Wiley, 1988.

\bibitem{3GPPTR36.859}
\BIBentryALTinterwordspacing
{3GPP TR 36.859 V0.1.0}, ``\textsc{3GPP}; technical specification group radio
  access network; study on downlink multiuser superposition transmission
  (\textsc{MUST}) for \textsc{LTE},'' 2015. [Online]. Available:
  \url{\url{www.portal.3gpp.org}}
\BIBentrySTDinterwordspacing

\bibitem{Iwabuchi}
M.~Iwabuchi, A.~Benjebbour, Y.~Kishiyama, G.~Ren, C.~Tang, T.~Tian, L.~Gu,
  T.~Takada, and T.~Kashima, ``5g field experimental trials on urllc using new
  frame structure,'' in \emph{IEEE Globecom Workshops}, 2017.

\bibitem{Le2021}
T.-K. Le, U.~Salim, and F.~Kaltenberger, ``An overview of physical layer design
  for ultra-reliable low-latency communications in 3gpp releases 15, 16, and
  17,'' \emph{IEEE Access}, vol.~9, pp. 433--444, 2021.

\end{thebibliography}

\end{document}